\documentclass[4paper]{aa}

\usepackage[varg]{txfonts}

\defcitealias{Vazdekis10}{V10} 
\defcitealias{Vazdekis12}{V12} 
\defcitealias{Vazdekis03}{V03} 
\defcitealias{Roeck15}{R15}

\usepackage[countmax]{subfloat}
\usepackage{rotating}
\usepackage{natbib}
\usepackage{longtable}
\usepackage{hyperref}
\usepackage{graphicx}	% Including figure files
\usepackage{amsmath}	% Advanced maths commands
\usepackage{amssymb}	% Extra maths symbols
\usepackage{textcomp}
\bibpunct{(}{)}{;}{a}{}{,}

\usepackage[T1]{fontenc}
\usepackage{ae,aecompl}

\begin{document}

\title {MILES extended: Stellar population synthesis models from the optical to the infrared}
\author {B. R\"{o}ck\inst{1,2} 
   \and A. Vazdekis \inst{1,2} 
  \and  E. Ricciardelli \inst{3} 
  \and  R.F. Peletier \inst{4} 
  \and  J.H. Knapen \inst{1,2} 
  \and  J. Falc\'{o}n-Barroso \inst{1,2} }

\institute{Instituto de Astrof\'{i}sica de Canarias, V\'{i}a L\'{a}ctea s/n, E-38205 La Laguna, Tenerife, Spain
\and Departamento de Astrof\'{i}sica, Universidad de La Laguna, E-38205 La Laguna, Tenerife, Spain
\and Departament d'Astronomia i Astrofisica, Universitat de Valencia, c/Dr. Moliner 50, E-46100 Burjassot, Valencia, Spain
\and Kapteyn Astronomical Institute, University of Groningen, Postbus 800, 9700 AV Groningen, The Netherlands}

\date{}

\titlerunning{MILES extended: Stellar population synthesis models from the optical to the IR}

\abstract {We present the first single-burst stellar population models which covers the optical and the infrared wavelength range between 3500 and ${\rm 50000  \, \AA}$ and which are exclusively based on empirical stellar spectra. To obtain these joint models, we combined the extended MILES models in the optical with our new infrared models that are based on the IRTF (Infrared Telescope Facility) library. The latter are available only for a limited range in terms of both age and metallicity. Our combined single-burst stellar population models were calculated for ages larger than 1 Gyr, for metallicities between ${\rm [Fe/H] = -0.40}$ and 0.26, for initial mass functions of various types and slopes, and on the basis of two different sets of isochrones. They are available to the scientific community on the MILES web page. We checked the internal consistency of our models and compared their colour predictions to those of other models that are available in the literature. Optical and near infrared colours that are measured from our models are found to reproduce the colours well that were observed for various samples of early-type galaxies. Our models will enable a detailed analysis of the stellar populations of observed galaxies. }


\maketitle

\section{Introduction}

Single-burst stellar population (SSP) models mimic uniform stellar populations of fixed age and metallicity, and are an important tool to study unresolved stellar clusters and galaxies. They are created by populating theoretical stellar evolutionary tracks with stars of a stellar library, according to a prescription given by a chosen initial mass function (IMF). Thus, the quality of the resulting SSP models depends significantly on the completeness of the used input stellar library in terms of evolutionary phases represented by the atmospheric parameters temperature, $T{\rm _{eff}}$, surface gravity, ${\rm log(g)}$, and metallicity. A sufficiently large spectral coverage is equally crucial when constructing reasonable SSP models. Theoretical stellar libraries like, e.g. BaSeL \citep{Kurucz92, Lejeune97, Lejeune98, Westera02}, or PHOENIX \citep{Allard12, Husser13} are generally available for both a large range in wavelength and in stellar parameters, whereas empirical libraries are found to be more incomplete in both respects. However, the advantage of the latter ones is that they are not hampered by the still large uncertainties in the calculation of model atmospheres. Examples of empirical stellar libraries in the optical wavelenth range encompass the Pickles library \citep{Pickles98}, ELODIE \citep{Prugniel01}, STELIB \citep{LeBorgne03}, Indo-US \citep{Valdes04}, MILES \citep{Sanchez06}, and CaT \citep{Cenarro01, Cenarro07}. In the near-infrared (NIR) and mid-infrared (mid-IR)\footnote{Note that throughout this paper, we refer to the wavelength range until the end of the $K$-band as NIR, and the range between 2.5 and ${\rm 5 \, \mu m}$ as mid-IR.}, only very few empirical libraries have been observed so far \citep[e.g.][]{Lancon00, Cushing05, Rayner09}. The NASA Infrared Telescope Facility (IRTF) spectral library, described in the latter two papers, is to date the only empirical stellar library in the NIR and mid-IR which offers a sufficiently complete coverage of the stellar atmospheric parameter space to construct SSP models. In the future, the X-Shooter stellar library, which contains around 700 stars, and which covers the whole optical \citep[see][]{Chen14} and NIR wavelength range until ${\rm 2.5 \, \mu m}$, will clearly improve the current situation in the NIR. 

In the optical range, many different sets of SSP models have been developed, based on the abovementioned libraries \citep[e.g.][]{Worthey94, Vazdekis99, Bruzual03, LeBorgne04, Maraston05, Maraston09, Conroy09, Vazdekis10}. Other models cover the NIR and mid-IR ranges, like those by \citet{Maraston05}, \citet{Conroy12}, \citet{Meneses15b}, and \citet[][hereafter R15]{Roeck15}.

%All these models differ mainly in their treatment of the so-called thermally pulsating asymptotic giant branch (TP-AGB) phase. Stars in this evolutionary stage are assumed to dominate the total luminosity of stellar populations of ages between 0.5 and 1.5 Gyr \citep[e.g.,][]{Maraston05, Marigo08}. However, their exact contribution is still highly debated. While, e.g., \citet{Maraston05, Maraston09, Maraston11} favour a much enhanced contribution of AGB stars to their models in order to be able to correctly explain observed IR colours, \citet{Zibetti13} did not find observational evidence for such an important impact of this kind of stars from a sample of 16 post-starburst galaxies.

In the current work, we combine our IRTF-based SSP models in the NIR and mid-IR with the MIUSCAT models \citep{Vazdekis12, Ricciardelli12} in the optical range. Thus we end up with combined SSP models, which are based on empirical stellar spectral libraries and encompass the spectral range from 3465 to ${\rm 50000 \, \AA}$. Moreover, we analyse and validate our new models in the NIR wavelength range between 1 and ${\rm 2.5 \, \mu m}$. 

%Due to the limitations of the IRTF library which contains only around 200 stars of mostly solar metallicities, our models are only available for solar abundance ratios, metallicities between ${\rm -0.40 < [Fe/H] < 0.20}$ and ages larger than 1 Gyr. So far, these are the only SSP models available in the literature for this entire spectral range which are -- apart from two small gap regions in the NIR and IR \citepalias[see Section 7 of][]{Roeck15}  -- completely based on various sets of empirical stellar spectral libraries. Apart from In our analysis, we focus on the NIR wavelength range which we have not yet publishedhave not yet discussed before and on combined NIR optical colours. 

%The paper is structured as follows. In Section \ref{models}, we recapitulate briefly the main ingredients and characteristics of both the MILES models in the optical \citepalias{Vazdekis12} and of the IRTF based models in the NIR and IR \citepalias{Roeck15}. After that, we describe how the joining between the two sets of models was carried out. Section \ref{Colour_analysis} is dedicated to a sanity check proving the internal consistency of our models as well as to a detailed study of the behaviour of various optical and NIR colours as a function of age and metallicity. In addition, we compare several colours measured from our models to those observed for early-type galaxies and globular clusters. Finally, we also compare the predictions of our models for some of the studied colours to those of other models in the literature. We end the paper with a summary (Section \ref{summary}).

\section{Models from the Optical to the IR}\label{models}

In this Section, we describe how we combined the MIUSCAT models \citep{Vazdekis12}, which range from 3465 to ${\rm 9469 \, \AA}$, with our newly developed SSP models in the NIR and mid-IR \citepalias{Roeck15}. We briefly recall the main ingredients and characteristics of the models in the two spectral ranges. A more detailed explanation of the modelling procedure is given in \citet{Vazdekis03}, \citet[][hereafter V10, V12]{Vazdekis10, Vazdekis12} for the optical and in \citetalias{Roeck15} for the IR part of the models. 
%All limits in the covered parameter space are imposed by the IRTF-based models which are based on only 180 stars \citepalias{Roeck15}. The other three libraries used for the modelling in the optical spectral range contain significantly more stars \citet{Cenarro01, Valdes04, Sanchez06}. 

\subsection{Main ingredients and characteristics}\label{models_ingredients}

Both the MIUSCAT models in the optical wavelength range and the IRTF-based models in the IR were calculated based on the isochrones of \citet{Girardi00} (hereafter: Padova00 isochrones) and on those of \citet{Pietrinferni04} (BaSTI isochrones). All of our used sets of isochrones are scaled-solar ones. 

It is important to note that the MIUSCAT models \citepalias{Vazdekis12} themselves are a compilation based on three different stellar libraries. The main part - from 3525 to 7410 \AA - consists of the MILES stellar library \citepalias{Vazdekis10}, whereas bluewards and redwards of it the Indo-U.S. spectral library \citep{Valdes04} and the CaT one \citep{Vazdekis03} were used in order to extend the total spectral range coverage. 

%The MILES stellar library \citep{Sanchez06} contains 985 stars covering the optical spectral range, while the CaT library \citep{Cenarro01} consists of 706 stars in the range of the Calcium triplet (8350 to ${\rm 9020 \, \AA}$). The INDO-U.S. library \citep{Valdes04} finally encompasses around 1200 stars covering typically the spectral range from 3465 to ${\rm 9469 \, \AA}$. 

%Consequently, the joined MIUSCAT model spectra range from 3465 up to 9469 \AA. Contrary to our approach when joining the MIUSCAT models with the IRTF-based ones (see next Section \ref{models_joining}), \citetalias{Vazdekis12} created combined stellar spectra taking advantage of stars in common between the three different stellar libraries and using an interpolator \citep{Vazdekis03, Vazdekis10} to synthesize stellar spectra not available in all of the libraries based on stars of similar atmospheric parameters. 

Whereas the MIUSCAT models are also available for young ages between 0.03 and ${\rm 1\, Gyr}$ (BaSTI isochrones) and 0.063 and ${\rm 1\, Gyr}$ (Padova00 isochrones), respectively, as well as for highly subsolar metallicities \citep{Vazdekis15}, our IRTF-based IR models presented in \citetalias{Roeck15} are of sufficient quality only for ages larger than ${\rm 1 \, Gyr}$ and for metallicities between ${\rm [Fe/H] = -0.40}$ and ${\rm \approx [Fe/H] = 0.25}$ due to the limited stellar atmospheric parameter coverage of the only 180 used stars of the IRTF library \citepalias[see Figure 5 in][] {Roeck15}, compared to the 985 stars of the MILES library \citep{Sanchez06}. For example, the IRTF library does not contain many young hot stars of $T{\rm_{eff} > 7000 \, K}$ and only a limited number of AGB stars which are important at ages smaller than ${\rm 1 \, Gyr}$ and between 1 and ${\rm 2 \, Gyr}$, respectively.  As described in \citetalias{Roeck15}, we determined the spectral resolution of the stars of the IRTF library to be ${\rm \sigma = 60 \, kms^{-1}}$ (which is equal to a resolving power of ${\rm R = \frac{\lambda}{\Delta(\lambda)}=\frac{c}{\sigma \cdot 2.35} \approx 2000 }$).

We calculated our SSP models for unimodal and bimodal IMFs of slopes between 0.3 and 3.3 \citep[for a more detailed description of the different IMFs see][] {Vazdekis96, Vazdekis03} as well as for the standard Kroupa and revised Kroupa IMFs \citep{Kroupa01}. %Since the influence of the IMF on the colours predicted by our models is almost negligible \citepalias{Roeck15}, we restrict our colour analysis to SSP models based on Kroupa IMFs.  

Moreover, our modelling code uses various photometric libraries and colour-temperature relations to convert the theoretical parameters ($T_{\text{eff}}$, $\log(g)$, [Fe/H]) to the observational plane. This approach is one key difference compared with most other sets of SSP models which generally take advantage of theoretical stellar atmospheres to perform this conversion. Primarily, our transformations are based on the metallicity-dependent relations of \citet{Alonso96, Alonso99} for dwarfs and giants, respectively. These relations were derived based on compilations of around 500 dwarf and giant stars, respectively, whose temperatures were determined using the IR-flux method. For the coolest dwarfs and giants of temperatures smaller than 3500 K as well as the hottest stars (${\rm T_{eff} > \approx 8000 \, K}$) the empirical compilation of \citet{Lejeune97, Lejeune98} was applied.

Table \ref{Parameter_coverage} summarizes the spectral properties and the parameter coverage of our combined SSP SEDs.

\begin{figure}
\begin{center}
 \resizebox{\hsize}{!}{\includegraphics{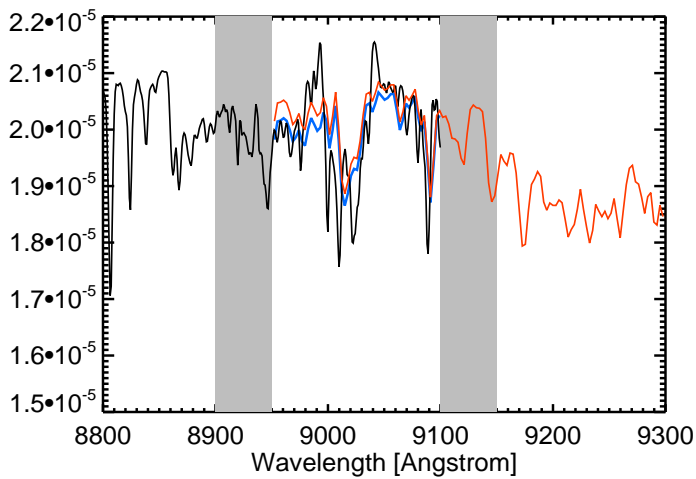}}
  \caption{Illustration of the combination of the MIUSCAT models (black line) with the IRTF based models (red solid line) between ${\rm 8950 \,\AA}$ and ${\rm 9100 \, \AA}$. The blue line shows the combined model, obtained by means of a slight adjustment which we performed on the IRTF- based model spectra in order to end up with the same continuum flux as the extended MILES model spectra. The continuum fluxes were measured within the two narrow grey-shaded bands ranging from 8900 to ${\rm 8950 \, \AA}$ and from 9100 to ${\rm 9150 \, \AA}$, respectively. Here, we plot model spectra of Solar metallicity and of an age of ${\rm 10 \, Gyr}$, based on Padova00 isochrones. Note the significantly higher resolution of the MIUSCAT model spectrum compared to the IRTF-based one.} \qquad
  \label{combination}
\end{center}
\end{figure}

\subsection{Combining the optical with the infrared part of the models}\label{models_joining}

In this work we take the calculated MIUSCAT \citepalias{Vazdekis12} model spectra and combine them with the IRTF model spectra of the corresponding parameters age, metallicity and IMF. Contrary to what was done when creating the MIUSCAT models by \citetalias{Vazdekis12}, we perform the joining directly between the two sets of models and not on a stellar level. For this purpose, we first measure the flux contained within the two narrow bands between 8900 and ${\rm 8950 \, \AA}$ and between 9100 and  ${\rm 9150 \, \AA}$ for both the MIUSCAT model spectra and for those based on the IRTF stellar library. These two bands were carefully selected in order not to encompass prominent absorption features but to be at the same time wide enough to achieve reasonable statistics. They are delineated in grey in Figure \ref{combination}. For both bands, we calculated the ratios between the fluxes from the two different sets of models. Then, we used the resulting values to rescale the flux of the IRTF-based model spectra linearly, pixel by pixel, between ${\rm 8950 \,\AA}$ and the beginning of the second window at ${\rm 9100 \, \AA}$. By using a correction which changes linearly with wavelength instead of a constant value throughout the whole joining region, we made sure not to introduce artificial jumps in the model spectra. The corrected IRTF spectra end up having the same continuum flux as the MILES spectra between 8900 and ${\rm 8950 \, \AA}$. The resulting combined SSP models consist of the MIUSCAT models for wavelengths smaller than ${\rm 8950 \, \AA} $, the corrected IRTF-based models between 8950 and ${\rm 9100 \, \AA} $ and the original IRTF-based models for wavelengths redder than ${\rm 9100 \, \AA} $. Figure \ref{combination} illustrates the MILES and IRTF parts of our spectra as well as the corrected IRTF part between 8950 and ${\rm 9100 \, \AA}$.

It is remarkable that the IRTF-based model spectra had to be modified only very little to end up with the same continuum flux as the MILES-based ones. Without introducing any correction, the continuum fluxes of the two sets of model spectra already agree within an accuracy of ${\rm <  2 \%}$ in the joining region. The reason for this is the excellent flux calibration of both the IRTF library and of the CaT library \citep{Cenarro01} which was used to account for the gap redwards of the wavelength range of the MILES stellar library. \citet{Conroy12} for example carried out a similar approach of using MILES spectra in the optical and IRTF spectra in the NIR as input for their models. To bridge the gap between 7400 and ${\rm 8150 \, \AA}$, they made use of synthetic spectra. They, however, encountered problems with the flux calibration and as a result the colours measured from their models are only accurate to around ${\rm \approx 5 \, \%}$.

We decided to use the IRTF-based models throughout the joining region since this reddest part of the MIUSCAT models, based on the Indo-U.S. library, is characterized by a limited flux-calibration quality and strong telluric residuals \citepalias[see][]{Vazdekis12}. The MILES SSP models have a constant FWHM resolution of approximately ${\rm 2.5 \, \AA}$, while the IRTF-based ones in the IR are calculated for a constant instrumental velocity dispersion of ${\rm \sigma = 60 \, km \, s^{-1}}$. Hence, in order to correctly perform the joining, in the region between 8950 and ${\rm 9100 \, \AA}$, we smoothed the MILES SSP spectra to a constant FWHM of ${\rm 4.2 \, \AA}$, which corresponds to a velocity dispersion of ${\rm \sigma = 60 \, km \, s^{-1}}$. However,we stress that we kept the MILES part of the models at its constant FWHM resolution of ${\rm \approx 2.5 \, \AA}$ and the IRTF part at its constant spectral resolution of ${\rm R \approx 2000}$ after joining the models. Thus, at a wavelength of ${\rm 8950 \, \AA}$,  there is a sudden change in resolution. 
%As we found out in \citetalias{Roeck15}, in the case of the stellar spectra of the IRTF library, it is not the FWHM but the instrumental velocity dispersion $\sigma$ which is kept constant. 
The IRTF-based models were rebinned to the same pixel size of ${\rm 0.9 \, \AA \, pixel^{-1}}$ as the MIUSCAT models. 
%Since in this work we merely joined two sets of already existing SSP models without actually repeating the modelling procedure, we refrain from giving the details of it in this work. The reader is referred to \citetalias{Vazdekis10, Vazdekis12} and \citet{Vazdekis15} for a detailed description of the SSP modelling in the optical wavelength range and to \citetalias{Roeck15} as far as the IR range is concerned. 

\begin{table*}
\caption{ Properties of our combined model SSP SEDs}
\label{Parameter_coverage}
\centering
\begin{tabular}{l c}

\hline
Spectral coverage\\
\hline
  Spectral range & ${\rm 0.35-5 \, \mu m}$ \\ 
  Spectral resolution & ${\rm \lambda < 8950 \, \AA: FWHM = 2.5 \, \AA, \lambda > 8950 \, \AA: \sigma=60 \,km \, s^{-1}}$\\
  Dispersion  & ${\rm 0.9 \, \AA \cdot pixel^{-1}}$ \\
  \hline 

Parameter coverage\\
\hline
  IMF type & unimodal, bimodal, Kroupa, Kroupa revised\\ 
  IMF-slope (for unimodal and bimodal) & 0.3-3.3 \\
  Isochrones  &  BaSTI \citep{Pietrinferni04}, Padova00 \citep{Girardi00}\\
  Metallicity range(BaSTI) & ${\rm -0.35  < [Fe/H] < 0.26 }$ \\
  Metallicity range (Padova00) & ${\rm -0.40  < [Fe/H] < 0.22 }$ \\
  Age range(BaSTI)         & ${\rm 1 \, Gyr < \textit{t} < 14 \,Gyr}$ \\
  Age range (Padova00)      & ${\rm 1 \, Gyr < \textit{t} < 17.78 \,Gyr}$\\
\hline
\end{tabular}
%\tablefoot{
%\medskip
\end{table*}

\section{Colours measured from our models}\label{Colour_analysis}

In this Section, we first check the internal consistency and correctness of our models by comparing several optical-NIR colours derived from them to the so-called photometric predictions for these colours (see Section \ref{colors_consistency}). The latter ones are based on the MILES models in the optical as well as on various colour-temperature relations. Moreover, we carry out a detailed comparison with the colours predicted by other models in the literature (Section \ref{literature_compare}). Finally, we show the ability of our models to fit various optical and NIR colours observed for early-type galaxies (ETGs,see Section \ref{Colours_observations}).

In order to obtain colours from our SSP model spectra, we multiply them by the transmission of the respective filters and integrate. In this procedure, we calibrate the resulting magnitudes based on the Vega-spectrum of \citet{Castelli94} as a zeropoint. We have to keep in mind that according to \citet{Ricciardelli12} the colours of Vega are assumed to be $ (V-R) {\rm = 0 \, mag}$ and $(V-I){\rm  = -0.04 \, mag}$ in the Johnson-Cousins filters. Thus, we have to correct our obtained colours according to the latter relation.

In this work we are particularly interested in optical - NIR colours, i.e., colours combining filters from the optical and the NIR range of our joined SSP model spectra. Consequently, we calculated the magnitudes in the Buser $V$ \citep{Buser78}, Cousins $R$ and $I$ and the Johnson $J$ and $K$-bands \citep{Bessell79}. 
%Apart from that, the literature does not provide observations of globular clusters and galaxies in the $L$-band which we could compare with our models.
%We did not make use of the Johnson $H$-band, since we could not find a unique definition of it in the literature. 
Since the influence of the IMF on the colours predicted by our models is almost negligible \citepalias{Roeck15}, we restrict our colour analysis to SSP models based on Kroupa IMFs.

\subsection{Checking the internal consistency of the model colours}\label{colors_consistency}

\begin{figure}
\begin{center}
 \resizebox{\hsize}{!}{\includegraphics{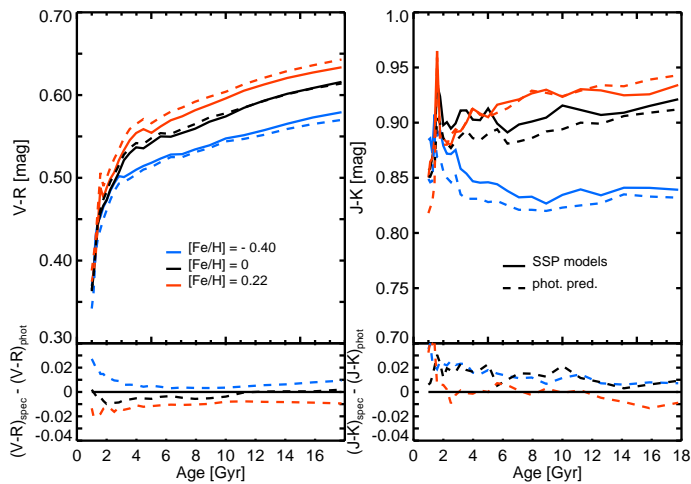}}
  \caption{Upper panels: comparison between the behaviour with age of the ($V-R$) and the ($J-K$) colours extracted from our  model spectra (based on Padova00 isochrones) and the same colours from the MILES photometric predictions \citep[see http://miles.iac.es][]{Vazdekis10, Vazdekis15}. The dotted lines delineate the MILES photometric predictions and the solid ones the colours calculated from our SSP model spectra. The different colours and symbols represent three different metallicites (see legend). Lower panels: Visualization of the absolute differences between the colours measured from our model spectra and the photometrically predicted ones. }\qquad
  \label{VR_JK_versus_age}
\end{center}
\end{figure}

\begin{figure}
\begin{center}
 \resizebox{\hsize}{!}{\includegraphics{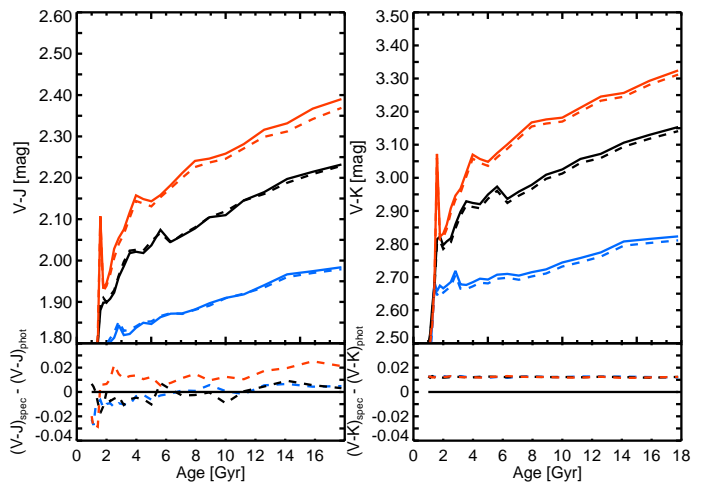}}
  \caption{Same as Figure \ref{VR_JK_versus_age}, but for ($V-J$) and ($V-K$) colours. }\qquad
  \label{VJ_VK_versus_age}
\end{center}
\end{figure}

\begin{figure}
\begin{center}
 \resizebox{\hsize}{!}{\includegraphics{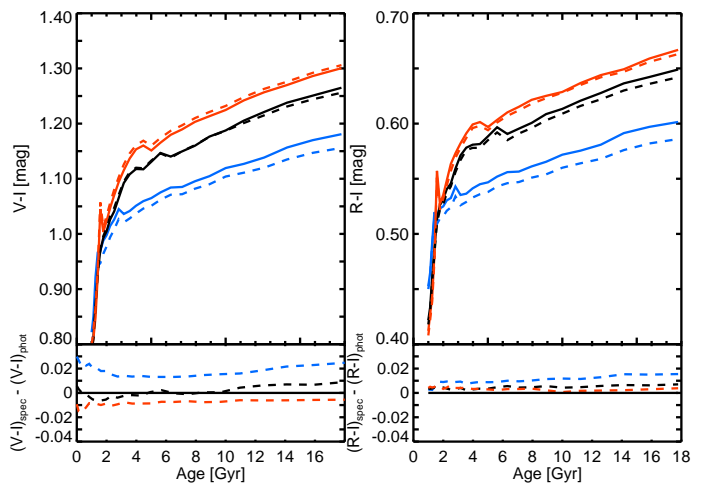}}
  \caption{Same as Figure \ref{VR_JK_versus_age}, but for ($V-I$) and ($R-I$) colours. }\qquad
  \label{VI_RI_versus_age}
\end{center}
\end{figure}

\begin{figure}
\begin{center}
 \resizebox{\hsize}{!}{\includegraphics{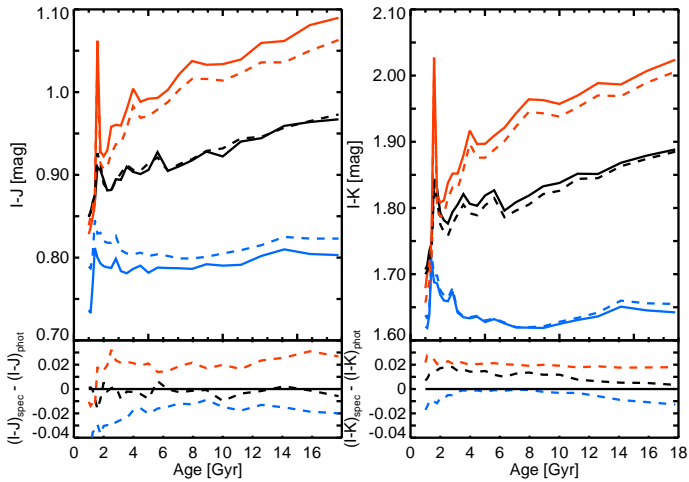}}
  \caption{Same as Figure \ref{VR_JK_versus_age}, but for ($I-J$) and ($I-K$) colours. }\qquad
  \label{IJ_IK_versus_age}
\end{center}
\end{figure}

In Figures \ref{VR_JK_versus_age} - \ref{IJ_IK_versus_age}, we compare various optical - NIR colours as calculated from our model SSPs with the same colours from the MILES photometric predictions. Since these photometric predictions have only been published based on Padova00 (see http://miles.iac.es), we present only comparisons to models using this set of isochrones. We also compared, however, the same model colours to the photometric predictions based on BaSTI isochrones and obtained very similar results. The photometric colour predictions have been determined by integrating the stellar fluxes obtained from the colour-temperature relations determined for dwarfs \citep{Alonso96} and for giants \citep{Alonso99} of spectral types F0-K5. These relations are based on extensive photometric stellar libraries and are in general metallicity dependent. Note that these colours are not obtained by integrating any synthesized SSP model SEDs within the bandpass of a filter. Since we have also used the same colour-temperature relations in order to transform the theoretical parameters (T$_{eff}$, log(g), [Fe/H]) of our models to the observational plane (see Section \ref{models_ingredients}), we expect a close agreement between the colours  from our models and those originating from the photometric predictions.

As can be seen from the lower panels of Figures  \ref{VR_JK_versus_age} - \ref{IJ_IK_versus_age}, the differences between the colours ${\rm \Delta(colour) = (colour)_{SSP -models} - (colour)_{phot. predictions}}$ are indeed almost for all cases smaller than the photometric precision of 0.02 mag. Slightly larger differences can usually be observed for the young ages (${\rm < 2 \, Gyr}$), which is due to the lower quality of our models in this age range (see Section \ref{models}) caused by the limited number of AGB and carbon stars in the IRTF library. In the NIR, these types of stars contribute significantly to the total flux at these young ages \citep[e.g.][]{Maraston05, Maraston09, Maraston11}. Apart from that, the differences between the colours measured directly from the spectra and the ones from the photometric predictions tend to be the largest in the models with supersolar metallicity, ${\rm [Fe/H]=-0.22}$. These are also of lower quality compared to their counterparts of Solar and subsolar metallicity since the coverage in parameter space is rather poor for such metallicities in the IRTF library \citepalias[compare to Section 8.3 in][] {Roeck15}.

The good agreement which we find between the colours measured from our model spectra and the photometrically predicted ones confirms that we have carried out correctly the modelling process and that hence our models are internally consistent. It also shows the good flux calibration of the IRTF stellar library which we confirmed in \citetalias{Roeck15}. Moreover, also in the optical range, \citet{Ricciardelli12} found differences of the order of only ${\rm 0.01-0.02 \, mag}$ between the synthetic and the photometric colours. To test whether the joining of the optical with the infrared part of our models was done correctly, the colours containing the $I$ filter are particularly interesting, since that encompasses the joining region of the models from 8900 to ${\rm 9150 \, \AA}$. The same holds for colours composed of an optical and a NIR band like ($V-J$) and ($V-K$). The fact that these colours do not show a worse agreement with the photometric predictions than pure optical or NIR colours proves the robustness of our method of joining the MIUSCAT spectral range with the IRTF-based one described above (see Section \ref{models_joining}).

%\begin{figure}
%\begin{center}
% \resizebox{\hsize}{!}{\includegraphics{RJ_RK_versus_age_BB.eps}}
%  \caption{Same as Figure \ref{VR_JK_versus_age}, but for ($R-J$) and ($R-K$) colours. }\qquad
%  \label{RJ_RK_versus_age}
%\end{center}
%\end{figure}

\subsection{Comparison to other model predictions in the literature}\label{literature_compare}

We now compare the predictions of our models for colours to other models available in the literature which cover the optical and NIR wavelength range. These are in the first place the models by \citet[][hereafter Maraston models]{Maraston98, Maraston05, Maraston09} and the ones by \citet[][hereafter CvD models]{Conroy12}. Furthermore, the Padova group has presented colour predictions in this wavelength range based on their library of stellar isochrones \citep{Marigo08}. Recently \citet{Meneses15b} published SSP models in the NIR between 9300 and ${\rm 24100 \, \AA}$ which are also based on the stars of the IRTF library. In what follows, we summarize the main characteristics of all these models.

\subsubsection*{Maraston models}

The Maraston models are calculated for the sets of isochrones developed by \citet{Cassisi97a, Cassisi97b} and \citet{Cassisi00} and use as input the theoretical stellar spectra of the BaSeL stellar library \citep{Lejeune98}. The spectra of the BaSeL library are based on stellar atmospheres from \citet{Kurucz92} and, in the case of M giants, from \citet{Fluks94} and \citet{Bessell89, Bessell91}. For TP-AGB stars, \citet{Maraston05} include the empirical spectra of the library of \citet{Lancon02}. 

\subsubsection*{Model predictions by the Padova group}

\citet[][hereafter Marigo models]{Marigo08} also make use of the Kurucz stellar atmospheres. More in detail, they work with the revised ATLAS9 model spectra by \citet{Castelli03}, while adopting the model atmospheres of \citet{Allard00} for M, L and T dwarfs and those of \citet{Fluks94} for M giants. Contrary to the Maraston models, the Marigo models employ the synthetic spectra of \citet{Loidl01} for TP-AGB stars. 
%\citet{Bressan12} updated the Marigo models by incorporating a new reference Solar composition and a different treatment of convection. They also modified the way TP-AGB stars are treated \citep[see][]{Marigo13}, but so far have not included this evolutionary phase into their isochrones. 
The main caveat of the Marigo models is that they give only colour predictions and do not provide model spectra.

\begin{figure}
\begin{center}
 \resizebox{\hsize}{!}{\includegraphics{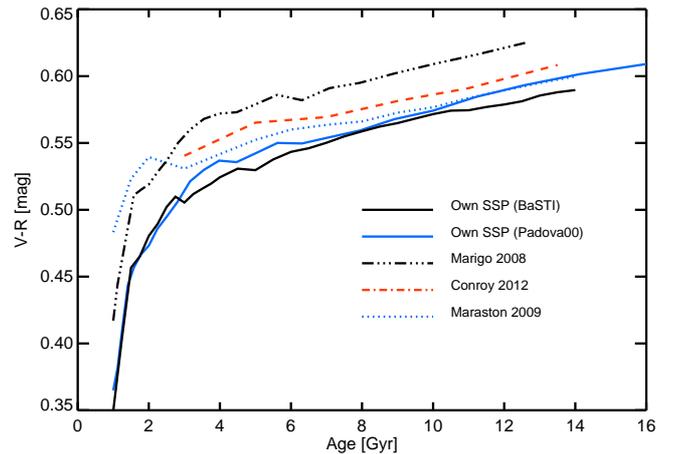}}
  \caption{Comparison of the behaviour of the ($V-R$) colour as a function of age for three different models (see legend) and our own (BaSTI- and Padova00-based) ones. All the displayed models are of solar metallicity and based on an underlying Kroupa IMF.}\qquad
   \label{VR_other_models}
\end{center}
\end{figure}

\subsubsection*{CvD models}

\citet{Conroy12}, however, followed an approach comparable to ours by combining stellar spectra of the two empirical stellar libraries MILES \citep{Sanchez06} and IRTF \citep{Rayner09} with similar parameters. The MILES spectra cover the spectral range between 0.35 and ${\rm 0.74 \, \mu m}$, whereas the IRTF stars are available for wavelengths between 0.81 and ${\rm 2.4 \, \mu m}$. In order to fill in the gap between the two stellar libraries and also in order to incorporate the possibility to change individual element abundances, \citet{Conroy12} used the ATLAS model atmospheres \citep{Kurucz92}. It should be noticed that \citet{Conroy12} used a subsample of the IRTF library only about half the size of ours (91 stars) by placing quite strong constraints on the individual stars (only luminosity classes III or V, measured parallaxes available). Moreover, as mentioned in Section \ref{models_joining}, colours measured from their models are less accurate than colours obtained from our models. To reproduce the main sequence and red giant branch evolutionary phases, \citet{Conroy12} applied the Dartmouth isochrones \citep{Dotter08}. This set of isochrones, however, does not exist for evolutionary phases beyond the core helium ignition. Hence, they used the Padova isochrones \citep{Marigo08} for the horizontal and asymptotic branches.

\subsubsection*{Discussion}

Fig. \ref{VR_other_models} shows the predicted ($V-R$) colours extracted from both our Padova00- and BaSTI-based SSP models, as well as from the three sets of models discussed above (see legend). We see that in ($V-R$), all of the models from the literature follow the same trend as our SSP models and become redder with increasing age. The largest change in colour can be observed for ages between 1 and ${\rm 3 \, Gyr}$, while the reddening is less abrupt for older ages. For the youngest ages, the Marigo models trace best the predictions from our SSP models. However, they exhibit the largest difference (${\rm \Delta \approx 0.03 - 0.05}$) for ages older than ${\rm 3 \, Gyr}$. In this age interval, the Maraston models and the Conroy ones agree closest with our models. Moreover, for these older ages, the predictions of our models show a good agreement with the observed colours of globular clusters, as can be seen from figures 1 and 3 in \citet{Ricciardelli12}.

\begin{figure}
\begin{center}
 \resizebox{\hsize}{!}{\includegraphics{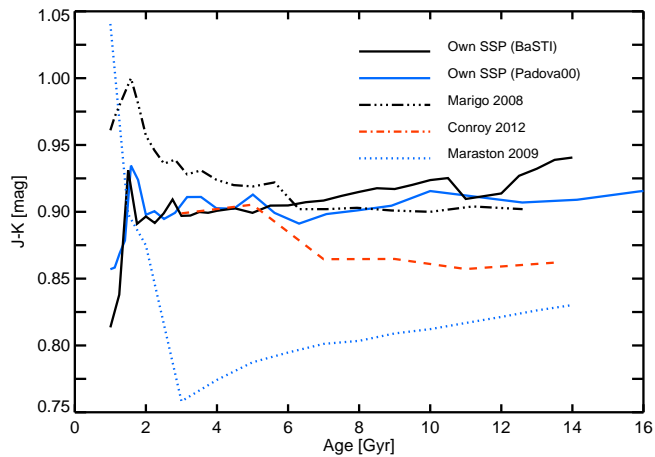}}
  \caption{Same as Fig. \ref{VR_other_models}, but for the ($J-K$) colour.}\qquad
   \label{JK_other_models}
\end{center}
\end{figure}

As Fig. \ref{JK_other_models} shows, the difference between the sets of models is much more pronounced for ($J-K$). Both of our SSP models show a moderate peak towards red colours for ages between 1 and ${\rm 2 \, Gyr}$, before becoming slightly redder again for older ages. We attribute this peak to the AGB and TP-AGB stars which are the main contributors to the integrated IR light at these intermediate ages. The same peak  -- though considerably stronger -- becomes visible in the Marigo models. For ages larger than ${\rm 3 \,Gyr}$, this set of models gives gradually slightly bluer colours. This behaviour, which is contrary to what we observe in our models, is also observed in the ($J-K$) colours of the CvD models. However, while the Marigo models coincide very well with ours for ages older than ${\rm 3 \, Gyr}$, the CvD models are too blue by about ${\rm \Delta}(J-K) \approx 0.06$ for ages larger than ${\rm 7 \, Gyr}$.  Due to an enhanced AGB star contribution \citep[see also][]{Maraston98, Maraston05, Maraston09}, the Maraston models show the reddest colours between 1 and ${\rm 2 \, Gyr}$ of all the models considered here. At ages of ${\rm 1 \, Gyr}$, the peak has apparently not yet been reached, and the reddening seems to continue for younger ages. Then, the ($J-K$) colour decreases rapidly until reaching a minimum at around ${\rm 3 \, Gyr}$. For older ages, the ($J-K$) colour becomes redder again, while always staying too blue by at least ${\rm \Delta}(J-K) \approx 0.07$ with respect to the predicted colours of our models.  

Recently, \citet{Zibetti13} has risen serious doubts about the much enhanced contribution of TP-AGB stars to the Maraston-models which results in the red colours for ages between 1 and 3 Gyr in Fig. \ref{JK_other_models}. \citet{Zibetti13} were unable to reproduce the observed optical-NIR colours of their 16 post-starburst galaxies with any combination of stellar populations based on the Maraston models. Despite the somewhat bluer colours predicted for older ages, both the assumption of a simple stellar population or a scenario of a recent burst of star formation together with an underlying old population yielded colours which were too red compared to their observations. 
%With our models, it is however possible to obtain very good fits to the observed colours of galaxies by assuming just one, rather old, single stellar population (see Fig. \ref{Frogel_1}). This result is consistent with the optical spectral range, where the observed colours can be also reproduced by an old, single stellar population without the need for a contribution of an additional young population. 

The comparison carried out here for the $(J-K)$ colour with our models and those of Marigo and CvD highlights the overly extreme behaviour of the Maraston models. In order to compensate the very red colours at younger ages caused by the large fraction of AGB stars contributing to their models, they end up with too blue colours at older ages. 

\subsubsection*{Models by \citet{Meneses15b}}

\begin{figure}
\begin{center}
 \resizebox{\hsize}{!}{\includegraphics{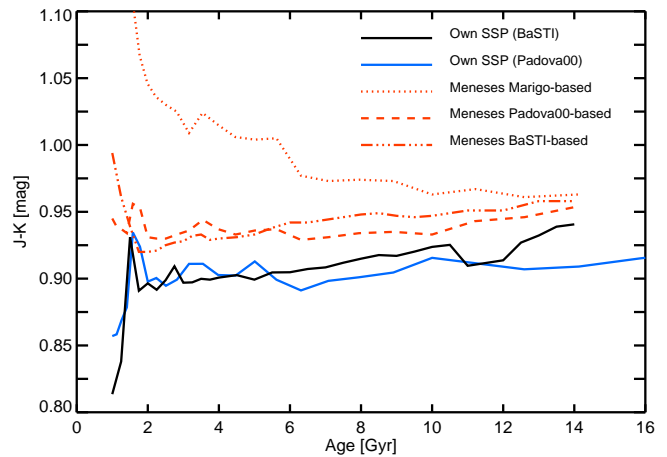}}
  \caption{As Fig. \ref{JK_other_models}, but comparing the $(J-K)$ colours of the three different sets of Meneses-models to our own (BaSTI- and Padova00-based) ones.}\qquad
   \label{JK_Meneses}
\end{center}
\end{figure}

To construct their models (hereafter Meneses-models), \citet{Meneses15b} excluded less stars from the IRTF library than we did and they also determined their atmospheric parameters differently. While we used both colour-temperature relations and a large literature compilation \citep{Roeck15}, \citet{Meneses15a} determined the stellar atmospheric parameters using full-spectrum fitting based on ULySS \citep{Koleva09}. Moreover, they slightly changed the interpolator code \citep{Vazdekis03}\footnote{These performed changes render a joining of their models with the extended MILES models in the optical \citepalias{Vazdekis10, Vazdekis12} impossible.} In addition to the BaSTI and the Padova00 isochrones, \citet{Meneses15b} computed their models also based on the set of isochrones by \citet{Marigo08}. 

Fig. \ref{JK_Meneses} shows a comparison between our model predictions for the $(J-K)$ colour and those of the different sets of Meneses-models. The colours originating from our models based on the BaSTI and the Padova00 isochrones agree well with those of the corresponding sets of the Meneses-models. Compared to our models, the latter ones are, however, generally redder by $\Delta (J-K) = 0.03 \, \rm{mag}$. This difference can be explained by the different adopted atmospheric parameters with respect to our models, in particular for the coolest stars. In addition, our models offer the advantage of a much larger wavelength coverage, since they include the whole optical wavelength range and extend up to ${\rm 5 \, \mu m}$.

The set of Meneses-models calculated using the isochrones of \citet{Marigo08} show a very similar overall behaviour to the colour predictions of the Marigo models (see Fig. \ref{JK_other_models}). The former yield, however, colours which are redder by about $\Delta (J-K) = 0.07 - 0.08 \, \rm{mag}$ and their increase due to the contribution of AGB stars at ages smaller than 2 Gyr is even more pronounced than in the Marigo models. Contrary to, e.g. the Maraston-models, the ($J-K$) colours originating from this set of Meneses-models also do not decrease significantly towards bluer colours for older ages. Moreover, the NIR-colours of the observed ETGs can be only fitted by models of low metallicities of ${\rm [Fe/H] = -0.4 - -0.7}$ (see Figure 10 in \citet{Meneses15b}). However, stellar population analysis in the optical range renders such low metallicities for ETGs extremely unlikely.

%\textbf{Therefore, we recommend the reader not to use the Meneses-models based on the Marigo-isochrones. Their other two sets of models based on BaSTI and Padova00 isochrones show a very similar behaviour to our models. However, our models offer the additional advantage of a much larger wavelength coverage, since they include the whole optical wavelength range and extend up to ${\rm 5 \, \mu m}$.}

\subsection{Reproducing observed colours of ETGs}\label{Colours_observations}

%Figures 9 -11 display a comparison between the 2MASS \citep{Skrutskie06} $J - K_s$, J-H and $H-K_{s}$ colours measured from our models with two different observational samples. The first one is a small subsample of seven out of the 54 massive star clusters in the Magellanic Clouds (MC) for which \citet{Pessev06, Pessev08} derived NIR  integrated-light magnitudes from the 2MASS catalogue. Since our SSP models do not reach the low metallicities of most of these star clusters, we were able to use only the seven ones of highest metallicity and youngest ages. The ages and metallicities which \citet{Pessev08} assumed for their MC star clusters originate from various literature sources and were in general determined using deep colour-magnitude diagrammes accompagnied by medium- and high-resolution spectroscopy of individual giant cluster stars. The second sample contains eleven early-type galaxies in the Fornax cluster \citep{Gonzales04} 

\begin{figure}
\begin{center}
 \resizebox{\hsize}{!}{\includegraphics{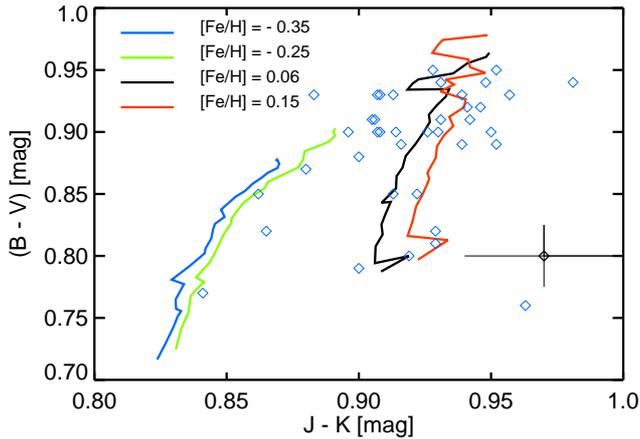}}
  \caption{($B-V$) versus ($J-K$) diagram showing our models overplotted on a sample of ETGs (blue diamonds). The uncertainties in both colours are indicated by an error bar. We display the predictions of our models based on BaSTI isochrones for ages between 1 and ${\rm 14 \, Gyr}$ and various metallicities (see legend).}\qquad
  \label{BV_JK_ETGs}
\end{center}
\end{figure}

Fig. \ref{BV_JK_ETGs} shows the optical colour ($B-V$) as a function of the NIR colour ($J-K$) for a selected sample of ETGs together with the predictions from our models. The 2MASS ($J-K$) colours were taken from the 2MASS Large Galaxy Atlas \citep{Jarrett03} while the corresponding optical ($B-V$) colours come from the HyperLeda\footnote{http://leda.univ-lyon1.fr/} database \citep{Makarov14}. The colours of our models fit very well those of the observed galaxies. This proves that our combined models qualify perfectly to reproduce all the colours from the optical until the mid-IR range.

\begin{figure}
\begin{center}
 \resizebox{\hsize}{!}{\includegraphics{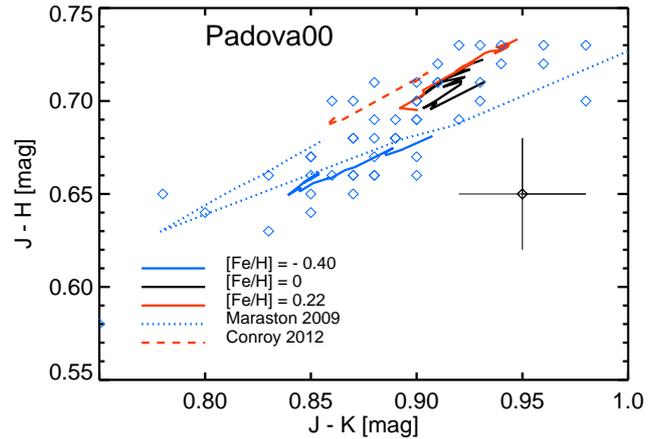}}
 \caption{($J-H$) versus ($J-K$) diagram of the ETGs of \citet{Frogel78} (blue diamonds). Overplotted are the corresponding colours predicted from our SSP models based on Padova00 isochrones (solid lines, colours indicate different metallicities according to the legend) as well as those originating from the Maraston models (blue dotted line) and from the CvD models (red dashed line). Both of these sets of models were computed for Solar metallicity and for ages larger than 1 Gyr (Maraston models) and 3 Gyr (CvD models), respectively. The observed uncertainties in both colours are indicated by an error bar. }\qquad
  \label{Frogel_1}
\end{center}
\end{figure}

In Fig. \ref{Frogel_1}, we compare the colour predictions of our models in the NIR  to the observed colours of the central regions of 51 bright ETGs (mostly ellipticals and S0) at small redshifts ($z < 0.022$) observed by \citet{Frogel78}. All of their colours are corrected to the restframe. Additionally, we also overplot the colour predictions of the Maraston and the CvD models for Solar metallicities, Kroupa-like IMFs and ages older than 1 Gyr and 3 Gyr, respectively. For both examined colours, ($J-K$) and ($J-H$), our models provide good fits. Like in the optical range, the observed colours can be reproduced well by an old, single stellar population without the need for an additional young component. 
%The best agreement is found in ($J-K$) and ($J-H$), whereas the scatter of the observed galaxies is slightly larger in ($H-K$). Nevertheless, 
Our models coincide with the mean colours of the observed sample in both colours. At the same time, the models also trace a substantial part of the scatter in the colours.  Both colours are hampered by the age-metallicity degeneracy, i.e. they become redder with both increasing age and metallicity. Consequently, the bluer galaxies can be both more metal-poor and younger than their redder counterparts. 

Both the colours obtained from the CvD models and from the Maraston ones show a worse agreement with the observations than the colours originating from our SSP models. While the CvD models yield somewhat too blue ($J-K$) colours compared to the bulk of ETGs, the Maraston models fit the observations reasonably well only for the very youngest ages between 1 and 2 Gyr. For ages older than 2 Gyr, the Maraston models provide colours which are all too blue with respect to the observed ETGs of \citet{Frogel78}. We noted this very extreme behaviour of the Maraston models already in Fig. \ref{JK_other_models} in Section \ref{literature_compare}.
% where we attributed it to their preferred combined scenario for ETGs consisting of a young, AGB-enhanced population on top of on underlying older one. 
Contrary to the Maraston models, our models allow us to fit the colours of the observed galaxies with a single SSP without the need for a contribution of an additional young population, which is in good agreement with detailed studies in the optical range.

\section{Summary}\label{summary}

We present the first evolutionary stellar population models covering the wavelength range from 3465 to ${\rm 50000  \, \AA}$ which are -- apart from two gaps which were covered with the help of theoretical libraries -- based on empirical stellar spectra. We obtained this large wavelength coverage by combining the extended MILES models \citepalias{Vazdekis10, Vazdekis12} in the optical range with the models based on the IRTF library \citep{Cushing05, Rayner09} in the NIR and mid-IR between 8150 and ${\rm 50000 \, \AA}$ \citepalias{Roeck15}. 

The actual matching was carried out between 8950 and ${\rm 9100 \, \AA}$ (see Section \ref{models_joining}). Thanks to the excellent flux calibration of the stellar spectra of both the MILES and the IRTF library, we had to adjust only slightly the fluxes in the joining region. Nevertheless, due to this slight artificial modification of the flux one should refrain from measuring line strength indices in this matching region. However, it is completely safe to obtain all kinds of optical and NIR/mid-IR colours by integrating the model spectra over the respective filter bandpasses. When determining line strength indices from our combined SSP models, one should be aware that the MILES-based part of the models is kept at the FWHM resolution of ${\rm 2.5 \, \AA}$, while the IRTF-based part redwards of ${\rm 8950 \, \AA}$ is given at the constant spectral resolution of ${\rm R \approx 2000}$. 
%which is equal to a instrumental velocity dispersion of ${\rm 60 \, km s^{-1}}$ \citepalias{Roeck15}. 

%Although we provide a good stellar population parameter coverage, our combined models are limited in parameter space by the constraints affecting the IR part (see Section \ref{models_ingredients}). Our IRTF-based SSP models make use of 180 stars of around Solar metallicity (${\rm -0.40 < [Fe/H] < 0.20}$). Due to an insufficient number of AGB stars and of young, hot stars, our models are not safe at ages younger than ${\rm 1 \, Gyr}$ \citepalias{Roeck15}. 
We calculated our models based on both BaSTI \citep{Pietrinferni04} and Padova00 \citep{Girardi00} isochrones. Our models are available for a large range of underlying IMFs, like Kroupa, revised Kroupa and unimodal and bimodal of slopes varying between ${\rm \Gamma = 0.3}$ and 3.3.
%Since we showed in Section 11.2 of \citet{Roeck15} that  colours derived from our models are almost independent of the IMF, we carry out our colour analysis (Section \ref{Colour_analysis}) exclusively for an underlying Kroupa-like IMF. Only when discussing line strength indices (see Section \ref{line_strengths_indices}), we take into account bimodal IMFs of different slopes. 

We conclude that our SSP models are internally consistent due to the close agreement of optical and NIR colours obtained from them with the colours originating from the predictions based on extensive photometric libraries published on our web page (http://miles.iac.es). Even colours composed of filters encompassing the joining region between our two sets of models usually do not deviate more than the photometric limit of ${\rm 0.02 \, mag}$ from the ones obtained from the photometric predictions. 

We also compared the predictions of our SSP models to those of other models which are available in the literature for the optical and NIR wavelength range. The observed differences between the theoretical Maraston, Marigo and the semiempirical CvD models and our two sets of Padova00 and BaSTI based SSP models are rather small in the case of the pure optical ($V - R$) colour. For a NIR colour like ($J - K$), some differences between the various sets of models arise from the different treatment of the TP-AGB evolutionary phase. Compared to our newly combined SSP models, the models of Marigo and in particular those of Maraston show a much more pronounced contribution of AGB stars which gets reflected in significantly redder colours with respect to our models for ages younger than 2 Gyr. For ages older than 3 Gyr, our SSP models agree best with the Marigo models, while the Maraston models show bluer colours compared to our models. The Meneses-models which are based on a comparable sample of stars from the IRTF library yield slightly redder colours in the case of underlying BaSTI andPadova00 isochrones. Their set of models calculated using the isochrones by \citet{Marigo08}, however, results in too red colours over the whole age range with respect to our models and in too low metallicities for the ETGs compared to what we know from the optical range. Moreover, our models offer the additional advantage of a much larger wavelength coverage.

Different optical and NIR colours measured from our SSP models are in good agreement with the observed colours from ETGs. Like in the optical spectral range, we are able to fit these galaxies with just one, old single-burst like stellar population and we do not need to assume an additional young component as is the case for other models in the literature. 

%Our models are also able to reproduce well the optical - NIR colours ($V - I$) and ($V - K{\rm_{s}}$) of globular clusters. 

Our combined models are available on the MILES webpage (http://miles.iac.es). Future work will focus on achieving a better understanding of the poorly analyzed absorption line features in the NIR from both a modeling and observational point of view.

\begin{acknowledgements}

We acknowledge financial support to the DAGAL network from the People Programme (Marie Curie Actions) of the European Union\textquotesingle s Seventh Framework Programme FP7/2007-2013/ under REA grant agreement number PITN-GA-2011-289313, and from the Spanish Ministry of Economy and Competitiveness (MINECO) under grant numbers AYA2013-41243-P and AYA2013-Y8226-C3-1-P. JHK thanks the Astrophysics Research Institute of Liverpool John Moores University for their hospitality, and the Spanish Ministry of Education, Culture and Sports for financial support of his visit there, through grant number PR2015-00512.

\end{acknowledgements}

\bibliographystyle{aa}

\bibliography{bibliography}

\end{document}